\documentclass[traditabstract]{aa} % for the abstract without structuration 
\usepackage{graphicx}
\usepackage{natbib}
\begin{document}
\title{
NGC\,6778: a disrupted planetary nebula around a binary central star\thanks{
Based on observations made with the Nordic Optical Telescope (NOT) and 
the Italian Telescopio Nazionale Galileo (TNG) on the island of La Palma 
in the Spanish Observatorio del Roque de los Muchachos of the Instituto 
de Astrof\'\i sica de Canarias (IAC).  
NOT is operated jointly by Denmark, Finland, Iceland, Norway, and Sweden. 
TNG is operated by the Fundaci\'on Galileo Galilei of the INAF 
(Istituto Nazionale di Astrofisica).  
The data presented here were obtained in part with ALFOSC, which is 
provided by the Instituto de Astrof\'\i sica de Andaluc\'\i a (IAA) 
under a joint agreement with the University of Copenhagen and NOTSA.
% Based on observations obtained at the Nordic Optical Telescope (NOT) and 
% Telescopio Nazionale Galileo (TNG) of the Observatorio de El Roque de los 
% Muchachos (ORM), and at the 4-m Blanco Telescope of the Cerro-Tololo 
% Inter-American Observatory (CTIO). 
}
}

%   \subtitle{I. Overviewing the $\kappa$-mechanism}

   \author{M.A. Guerrero
          \inst{1}\thanks{
Visiting astronomer, Cerro Tololo Inter-American Observatory, National 
Optical Astronomy Observatory, which are operated by the Association of 
Universities for Research in Astronomy, under contract with the National 
Science Foundation.}
          \and
          L.F.\ Miranda
          \inst{2,3}  % \fnmsep\thanks{Just to show the usage
                      % of the elements in the author field}
          }

\institute{Instituto de Astrof\'\i sica de Andaluc\'\i a (IAA-CSIC), 
           Glorieta de la Astronom\'\i a, s/n, E-18008 Granada, Spain 
\and
           Consejo Superior de Investigaciones Cient\'{\i}ficas, c/ 
           Serrano, 117, E-28006 Madrid, Spain 
\and 
Departamento de F\'\i sica Aplicada, Facultade de Ciencias, 
Campus Lagos-Marcosende s/n, Universidade de Vigo, E-36310 Vigo, Spain 
(present address)  \\
\email{mar@iaa.es,lfm@iaa.es}
}

\date{Received ; accepted }

% \abstract{}{}{}{}{} 
% 5 {} token are mandatory
 
\abstract{
The planetary nebula (PN) NGC\,6778 harbors a binary central star with a 
short orbital period and displays two systems of fast collimated outflows. 
In order to assess the influence of the evolution through a common-envelope 
phase of the binary system of NGC\,6778 on its formation and shaping, we 
have used narrow-band images and high-dispersion long-slit spectra of the 
nebula to investigate its detailed morphology and kinematics. 
We find that the overall structure of NGC\,6778 can be described as a 
bipolar PN.  
The equatorial ring is highly disrupted and many radial features 
(filamentary wisps and cometary knots) also evidence strong 
dynamical effects.  
There are clear connections between the bipolar lobes and the fast 
collimated outflows: the collimated outflows seem to arise from 
bright knots at the tips of the bipolar lobes, whereas the 
kinematics of the bipolar lobes is distorted.  
We suggest that the interaction of the fast collimated outflows of NGC\,6778 
with its nebular envelope has resulted in the disruption of the nebular shell 
and equatorial ring.  
%
% with its nebular envelope has transferred the kinetic energy needed to break 
% the symmetry of the nebular shell and shape the bipolar lobes.  
}


% \authorrunning{Guerrero and Miranda}
% \titlerunning{NGC\,6778: the planetary nebula around a binary central star}

   \maketitle
%
%________________________________________________________________

\section{Introduction}

Planetary nebulae (PNe) consist of stellar material ejected by 
stars with initial masses $\le$ 8--10 $M_\odot$.  
As such a star evolves off the asymptotic giant branch (AGB), it 
experiences copious mass loss episodes that dramatically reduce 
the stellar envelope.  
When the stellar core is finally exposed, the stellar effective temperature 
rises and, by the time it exceeds 30,000~K, the strong stellar UV radiation 
ionizes the circumstellar material and a new PN is born.

Our canonical view of the formation of PNe is based on the interacting 
stellar winds model on which the nebular shape is mostly determined by 
the interaction of the current fast, tenuous stellar wind of its central 
star \citep[e.g.,][]{PP91} with the previous slow, dense wind of 
the AGB phase \citep{Kwok83}.  
This model accounts for the gross morphology of PNe: a spherical symmetric 
AGB wind will result in a spherical PN, while an azimuthal density gradient 
in the AGB wind will result in an elliptical or bipolar PN, depending on the 
degree of the density gradient \citep{Balick87}.   
The observations of PNe have shown that their morphologies and detailed 
structures are extremely rich, including small-scale nebular features 
\citep[e.g., NGC\,7662;][]{Perinotto_etal04}, collimated bipolar outflows 
\citep[e.g., He\,2-90;][]{SN00,Guerrero_etal01}, and point-symmetric 
bubbles \citep[e.g., Hen\,2-47 and M\,1-37;][]{Sahai00} and collimated 
outflows \citep[e.g., NGC\,6884 and Hen\,3-1475;][]{
Borkowski_etal97,Miranda_etal99a,Riera_etal03}. 
This large body of observations reinforces the idea that the interacting 
stellar winds model needs to incorporate additional physical processes 
\citep[][]{BF02}.

Among the new physical processes to take into account, fast collimated 
outflows (jets) or collimated fast stellar winds can be expected to have 
profound effects in the shaping of PNe \citep{ST98}.
High-velocity bipolar jets impinging on the nebular material will 
transfer momentum and deposit kinetic energy into the nebula and 
break the spherical symmetry of the AGB wind.  
The prolonged action of collimated outflows may bore through the AGB wind 
and form extended cavities \citep[e.g., OH\,231.8+4.2;][]{Bujarrabal_etal02} 
whose orientation may change with the precession or rotation of the 
collimated outflow \citep[e.g., NGC\,6881;][]{GM98,KS05}.  
The strong shock of a fast collimated outflow may have dramatic effects 
in the nebula itself, causing the complete disruption of the nebular 
shell.

The mechanisms that generate fast collimated outflows in PNe is still an 
unsolved question \citep{Canto_etal88,FBL96,GarciaS_etal05,Blackman09}.  
Accretion disks formed during the common-envelope (CE) phase in a binary 
system have been proposed to be able to collimate fast outflows in PNe 
\citep{Morris87,Livio00,Soker04}, but compelling evidence of the 
association between close-binary nuclei and extremely asymmetric nebular 
morphologies and/or collimated outflows has been lacking \citep{BL90}. 
Deeper observations are nowadays strengthening this relationship 
\citep{Miszalski12}, and certainly the study of the detailed morphology 
and kinematics of PNe with binary central stars is of great interest.

Very recently, \citet{Miszalski_etal11} have revealed the binary 
nature of the central star of NGC\,6778 (PN\,G034.5$-$06.7).  
The short orbital period, $\sim$0.15 days, makes them conclude that 
it has undergone a CE phase.  
They also suggest the correlation between the binary evolution and the 
development of chaotic morphologies and fragmented shells and rings with 
wispy structures reminiscent of nova explosions.

\begin{figure*}
\centering
\includegraphics[width=6.5in]{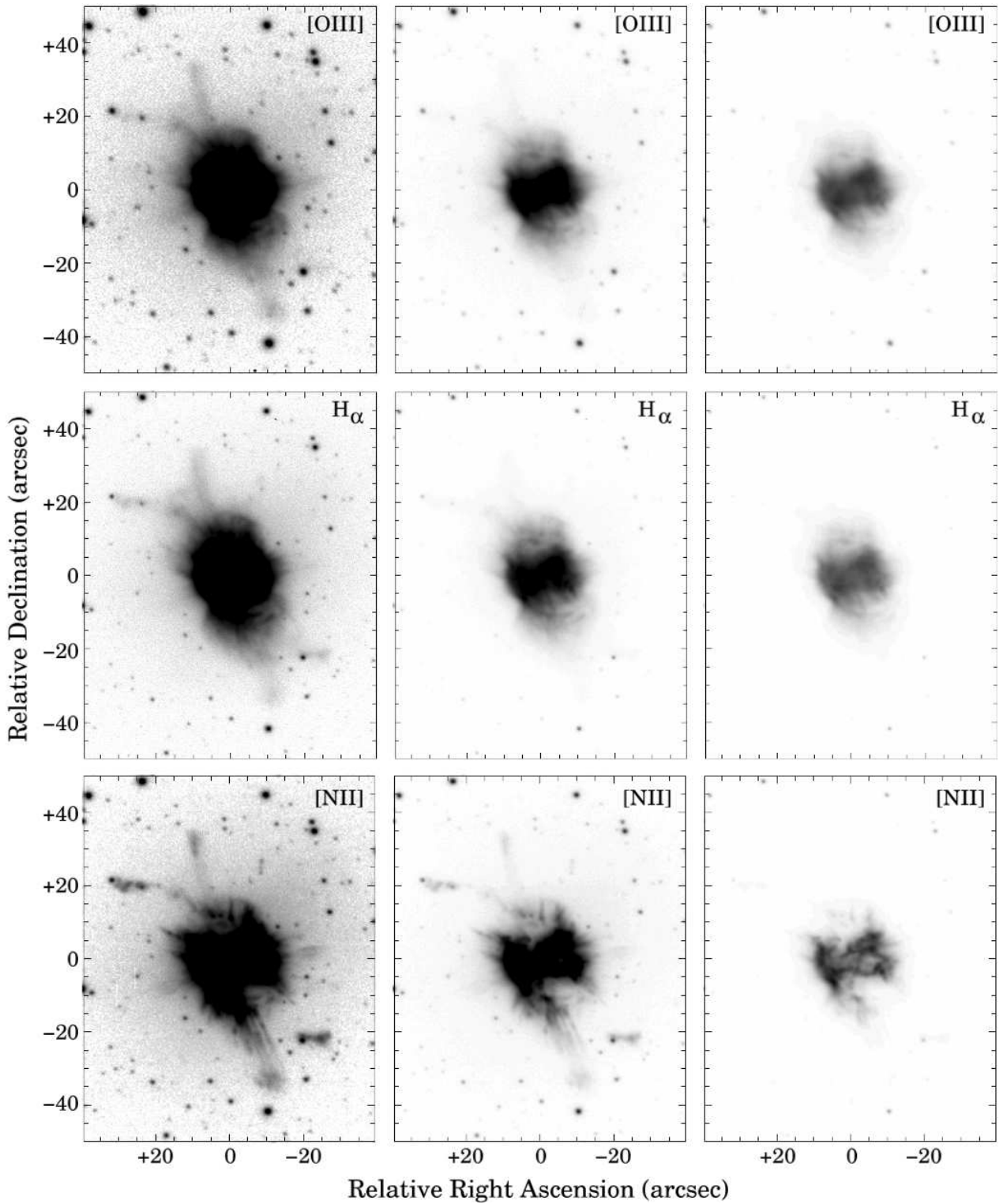}
\caption{
[O~{\sc iii}], H$\alpha$, and [N~{\sc ii}] narrow-band images of 
NGC\,6778. The images are shown with three different gray-scales to highlight 
the faint ({\it left}), medium ({\it center}), and bright ({\it right}) 
nebular features. The coordinate's origin corresponds to the position of the 
central star of the nebula, as detected in the near-IR images (see Fig.\,2).
}
\label{img1}
\end{figure*}

The morphology and kinematics of this PN had received little attention 
in the past.  
The first narrow-band H$\alpha$+[N~{\sc ii}] and [O~{\sc iii}] images 
of NGC\,6778 presented by \citet{SCM92} hinted one S-shaped feature 
protruding from an irregular shell.  
The deeper narrow-band H$\alpha$, [N~{\sc ii}], and [O~{\sc iii}] images 
presented by \citet[][hereafter Paper~I]{MGM04} confirmed and revealed greater 
details of this S-shaped feature, but also disclose an additional linear 
jet-like feature protruding from the central nebula along a different 
position angle.  
The morphology of these features is highly indicative of fast 
outflows, a suggestion that was confirmed by the radial
velocities of $\sim$100~km~s$^{-1}$ revealed by long-slit echelle 
spectroscopic observations (Paper~I).  
Due to projection effects, the expansion velocity of these collimated 
outflows is certainly greater than 100~km~s$^{-1}$.  
Notably, the radial velocity in the collimated outflows increases 
with the radius, as would be expected for the free expansion following 
a sudden episode of ejection.  
Higher resolution narrow-band images of NGC\,6778 described in the 
framework of a preliminary study of bipolar PNe at sub-arcsec scales 
\citep[][hereafter Paper~II]{MR-LG10}, revealed a bipolar structure 
of the nebular shell and a complex relationship between this shell 
and the fast collimated outflows.

In this paper, we revisit the kinematical data introduced in Paper~I 
in conjunction with the narrow-band images of higher spatial resolution 
of Paper~II to investigate the spatio-kinematical structure of the nebula, 
and to identify the radial velocity of individual morphological features.  
We describe the observations in Sect.\ 2 and analyze the morphology and 
kinematics of the nebula, and derive the near-IR properties of the 
central star in Sect.\ 3. 
The results are discussed in Sect.\ 4 and the conclusions are 
presented in Sect.\ 5.

\section{Observations}

\subsection{Narrow-band optical imaging}

Narrow-band images of NGC\,6778 have been acquired on 2005 August and 
2008 August using ALFOSC (Andalucia Faint Object Spectrograph and Camera) 
at the 2.56m Nordic Optical Telescope (NOT) of the Observatorio 
de Roque de los Muchachos (ORM, La Palma, Spain).  
The detector was an EEV 2048$\times$2048 CCD with a pixel size of 
13.5 $\mu$m, plate scale of 0$\farcs$19 pixel$^{-1}$, and field of 
view (FoV) 6$\farcm$5$\times$6$\farcm$5.  
The images were obtained through narrow-band filters that isolate the 
H$\alpha$ 
($\lambda_c = 6568$\,\AA, $\mathrm{FWHM} = 8$\,\AA), 
[N~{\sc ii}] $\lambda$6583 
($\lambda_c = 6589$\,\AA, $\mathrm{FWHM} = 9$\,\AA), and 
[O~{\sc iii}] $\lambda$5007 
($\lambda_c = 5007$\,\AA, $\mathrm{FWHM} = 8$\,\AA) emission lines.  
Table~\ref{img_obs} lists the number of individual frames acquired for 
each filter, their exposure time, and the total integration time.  
The data were bias-subtracted and flat-fielded by twilight flats using 
standard IRAF\footnote{
IRAF is distributed by the National Optical Astronomy Observatory, 
which is operated by the Association of Universities for Research 
in Astronomy, Inc., under cooperative agreement with the National 
Science Foundation.} 
routines.
We reproduce in Figure~\ref{img1} the [O~{\sc iii}] image of 
NGC\,6778 from the 2005 run, and the H$\alpha$ and [N~{\sc ii}] 
images from the 2008 run.  
The spatial resolution, as determined from the FWHM of stars in the FoV, 
was 0$\farcs$9 for the 2005 run and 0\farcs75 for the 2008 run.  
A preliminary description of these images was presented in Paper~II.

%
%_____________________________________________________________
%                                             Simple A&A Table
%_____________________________________________________________
%
\begin{table}
\caption{NOT narrow-band observations}
\label{img_obs}      
\centering           
\begin{tabular}{l l c c c}   
\hline\hline                 
\multicolumn{1}{l}{Date}             & 
\multicolumn{1}{l}{Filter}           & 
\multicolumn{1}{c}{Number of}        &
\multicolumn{1}{c}{Frame}            & 
\multicolumn{1}{c}{Total}            \\
\multicolumn{1}{c}{} & 
\multicolumn{1}{c}{} &
\multicolumn{1}{c}{frames} & 
\multicolumn{1}{c}{exp.\ time} & 
\multicolumn{1}{c}{exp.\ time} \\
\multicolumn{1}{c}{} & 
\multicolumn{1}{c}{} & 
\multicolumn{1}{c}{} & 
\multicolumn{1}{c}{(s)} & 
\multicolumn{1}{c}{(s)} \\
\hline             
August 2005 & H$\alpha$     & 2 & 300 & 600 \\
August 2005 & [N~{\sc ii}]  & 2 & 300 & 600 \\
August 2005 & [O~{\sc iii}] & 1 & 300 & 300 \\
August 2008 & H$\alpha$     & 2 & 450 & 900 \\
August 2008 & [N~{\sc ii}]  & 2 & 450 & 900 \\
\hline                                   
\end{tabular}
\end{table}

\subsection{Near-IR imaging}

\begin{figure*}
\centering
\includegraphics[width=1.9\columnwidth]{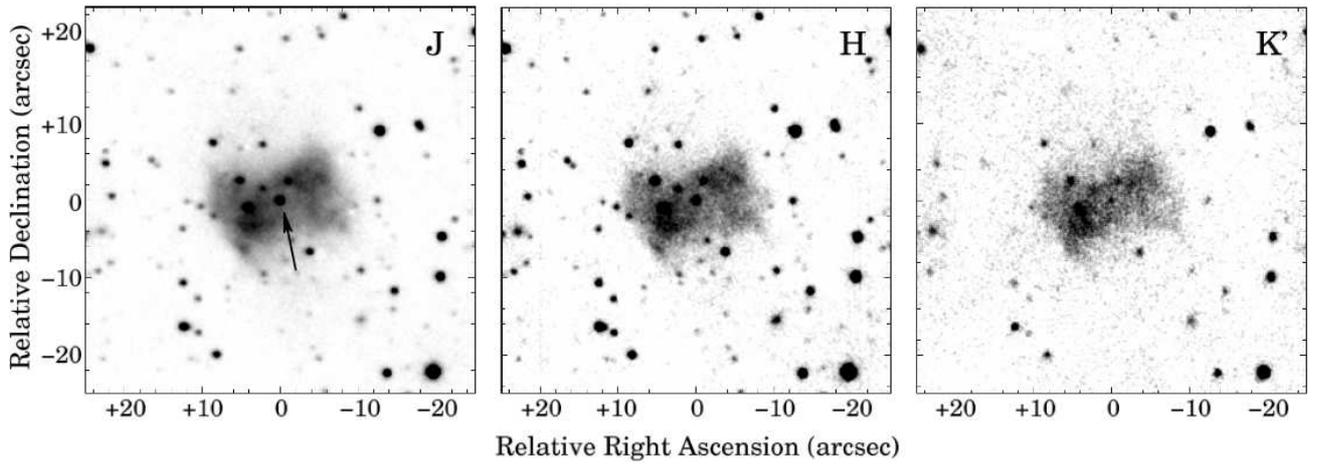}
\caption{
Grey scale representation of the near-IR broad-band $J$, $H$, and 
$K$ images of NGC\,6778. 
The grey levels are linear.  
The coordinate's origin corresponds to the position of the 
central star of the nebula detected in the three filters and 
marked by an arrow in the $J$ image.
}
\label{img_ir}
\end{figure*}

Near-IR images of NGC\,6778 were obtained on 2003 September 19 with
the Telescopio Nazionale Galileo (TNG) also at the ORM observatory.  
We used the NICS camera in the Small Field mode with a Rockwell 
HgCdTe 1024$\times$1024 array.  
The spatial scale on the detector is 0\farcs13 pixel$^{-1}$.  
Images were obtained through standard $J$, $H$, and $K^\prime$ filters 
with total exposure times of 1440 s in $J$ and $H$, and 360 s in $K^\prime$.  
The spatial resolution, determined by the FWHM of field stars, 
is $0\farcs70 - 0\farcs75$. 
The images were reduced following standard procedures within the 
MIDAS package and are shown in Figure~\ref{img_ir}.

\subsection{Optical long-slit high-dispersion spectroscopy}

High-dispersion optical spectra of NGC\,6778 were obtained on 2002 June 
23-24 using the echelle spectrograph on the 4m Blanco telescope of the 
Cerro Tololo Inter-American Observatory (CTIO).  
The spectrograph was used with the 79 line~mm$^{-1}$ echelle grating 
and the long-focus red camera, yielding a reciprocal dispersion of 
3.4 \AA~mm$^{-1}$. 
The SITE 2K \#6 CCD was used as detector.  
Its pixel size of 24~$\mu$m corresponds to a spatial scale of 
$0\farcs26\,\mathrm{pixel}^{-1}$ and a spectral scale of 
0.08\,\AA\,pixel$^{-1}$.  
The echelle spectrograph was used in single-order, long-slit mode by using 
a narrow-band filter that isolates the order covering the H$\alpha$ and 
[N~{\sc ii}] $\lambda\lambda6548,6583$ lines.  
The slit length was $\sim$3$^\prime$, while its width varied between 
0\farcs8 and 1\farcs2 for a spectral resolution $\sim$8\,km~s$^{-1}$.

\begin{figure}
\centering
\includegraphics[width=1.0\columnwidth]{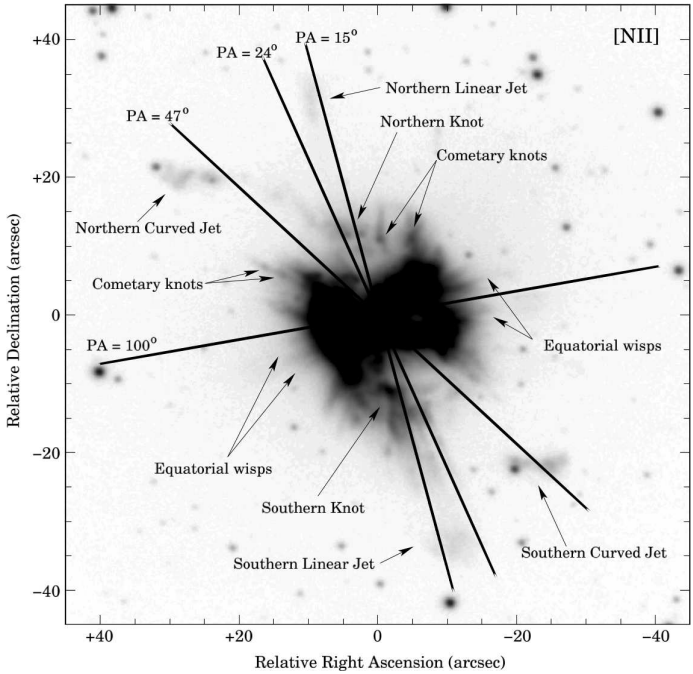}
\caption{
[N~{\sc ii}] image of NGC\,6778 overlaid by the positions of the 
slits used for the acquisition of the high-dispersion echelle 
spectroscopic observations at CTIO (slit width not to scale).  
Nebular features are also labeled on the figure. 
}
\label{img2}
\end{figure}

In order to register different nebular features of interest, the slit was 
oriented at position angles (P.A.) 15\degr, 24\degr, 47\degr, and 100\degr\ 
(measured following the standard convention east of north), as 
illustrated in Figure~\ref{img2}.  
The integration time of the spectra was 1200 s each.  
The spectra were wavelength calibrated with a Th-Ar arc lamp to an 
accuracy of $\pm$0.4\,km\,s$^{-1}$.  
The radial velocity of the source was measured relative to the systemic 
velocity, whence we deduce a value $V_{\rm LSR}$=106$\pm$2\,km\,s$^{-1}$ 
based on the average of the blue and red components of the radial velocity 
at the central position.  
This value implies a heliocentric velocity of 90$\pm$2\,km\,s$^{-1}$ 
which is in excellent agreement with the value of 91$\pm$3\,km\,s$^{-1}$ 
reported by \citet{DAZ98}.

\section{Results}

\subsection{The morphology of NGC\,6778}

\begin{figure}
\centering
\includegraphics[width=0.975\columnwidth]{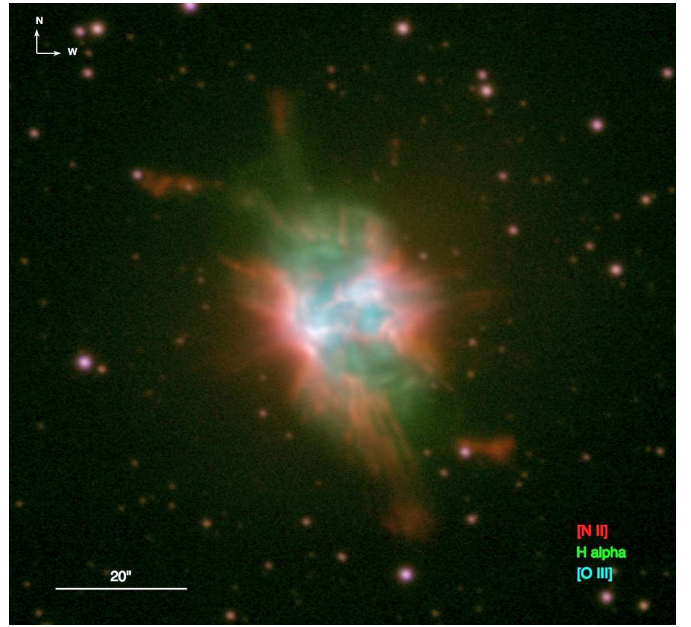}
\caption{
NOT composite color picture of NGC\,6778 in the [O~{\sc iii}] (blue), 
H$\alpha$ (green), and [N~{\sc ii}] (red) emission lines.  
}
\label{img3}
\end{figure}

The optical narrow-band images of NGC\,6778 (Fig.~\ref{img1} and \ref{img2}) 
provide a sharper view of the morphological components described by Paper~I 
and disclose new interesting features.  
Further information on the different morphological components 
of NGC\,6778 and their relative intensities in the H$\alpha$, 
[O~{\sc iii}], and [N~{\sc ii}] emission lines is provided by 
the composite-color picture shown in Figure~\ref{img3} and the 
[N~{\sc ii}]/[O~{\sc iii}] and H$\alpha$/[O~{\sc iii}] ratio 
maps presented in Figure~\ref{img4}.

\subsubsection{Main nebular shell}

The innermost region of NGC\,6778 can be depicted as a highly distorted 
and fragmented [N~{\sc ii}]-bright equatorial ring aligned along a 
line close to the east-west direction.  
Although H$\alpha$ and [O~{\sc iii}] emission is also found in these 
innermost regions, the spatial distribution of the emission from these 
lines is much smoother than that of [N~{\sc ii}].  
Note also that the values of the [N~{\sc ii}]/[O~{\sc iii}] and 
[N~{\sc ii}]/H$\alpha$ ratios peak at the location of this ring.  
By contrast, the emission in the H$\alpha$ and [O~{\sc iii}] 
lines is brighter along an axis at P.A.$\sim$20\degr, opening 
in bipolar lobes that span $\sim$17\arcsec.  
As suggested in Paper~I, the overall structure of the central region 
of NGC\,6778 can be described as a bipolar PN with a low-ionization 
ring-like structure embedded within higher excitation bipolar lobes.

\begin{figure}
\centering
\includegraphics[width=1.0\columnwidth]{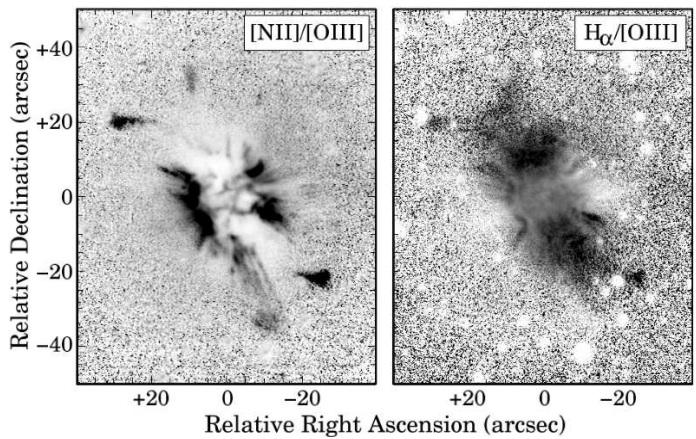}
\caption{
[N~{\sc ii}]/[O~{\sc iii}] and H$\alpha$/[O~{\sc iii}] ratio maps 
of NGC\,6778. 
Dark shades correspond to high ratio values.  
}
\label{img4}
\end{figure}

The composite-color picture in Figure~\ref{img3} and the ratio maps in 
Figure~\ref{img4} reveal clear excitation variations within the inner 
regions of NGC\,6778.  
At the tips of the equatorial ring, the H$\alpha$ and [O~{\sc iii}] 
emission is enclosed by bright [N~{\sc ii}] emission, resulting in 
the largest values of the [N~{\sc ii}]/[O~{\sc iii}] ratio.  
This [N~{\sc ii}]-bright structure encompasses a spheroidal 
region in the innermost section of NGC\,6778 with the lowest 
values in the H$\alpha$/[O~{\sc iii}] ratio (the bluish central 
region in the composite-color picture).  
This [O~{\sc iii}]-bright inner region is tipped by two regions of 
enhanced H$\alpha$/[O~{\sc iii}] ratios along the major axis of the 
bipolar lobes (seen in greenish colors in the composite-color 
picture).  
These changing ratios seem to imply that the nebula is highly 
ionization-bounded along the equatorial ring, while the bipolar 
lobes are density-bounded, although some shortage of ionizing 
photons at their tips is suggested by the enhanced H$\alpha$/[O~{\sc iii}] 
ratio.

The emission in the near-IR images of NGC\,6778 (Figure~\ref{img_ir}) 
is mostly limited to the innermost ring-like structure and to the 
[N~{\sc ii}]-bright regions at the tips of this ring.  
The emission from the bipolar lobes of NGC\,6778 is hinted in the $J$ 
band and, to less extent, in the $H$ band.  
Given the similarities between the H$\alpha$ and near-IR images 
of NGC\,6778, we suspect that the dominant contribution to the 
near-IR images are recombination lines of H~{\sc i}, in particular, 
the Pa$\beta$ $\lambda$1.282 $\mu$m in the $J$ band, the Brackett 
series in the $H$ band, and the Br$\gamma$ 2.166 $\mu$m in the $K$ 
band as seen in PNe whose near-IR spectra are dominated by H~{\sc i} 
lines \citep[][]{HLD99}.

\subsubsection{Linear and curved jet-like systems}

The NOT optical narrow-band images of NGC\,6778 reveal two jet-like 
systems, as labeled in Figure~\ref{img2}.  
These are detected in the three emission lines, but most notably in 
the [N~{\sc ii}] images because they are tipped by bright [N~{\sc ii}] 
emission caps.  
We name these two features as the linear and curved jet-like 
systems.  
The linear jet-like system (LJ) goes along P.A.$\sim$18\degr.  
Its northern component (NLJ) is mostly linear, but it shows a subtle tilt 
towards lower P.A.\ values by the time its [N~{\sc ii}] emission brightens 
at radial distances between $\sim$30\arcsec\ and 36\farcs5.  
The [N~{\sc ii}] image suggests that it originates 
from a knot marked as ``Northern Knot'' in Figure~\ref{img2}.  
The southern component of the linear jet (SLJ) consists of a series 
of filaments that, arising from a bright knot at a radial distance 
$\sim$10\arcsec\ (labeled as ``Southern Knot'' in Figure~\ref{img2}),  
run in parallel up to a distance of $\sim$30\arcsec, when they 
brighten into a 3\arcsec$\times$4\arcsec\ blob.  
Given our spatial resolution, it is unclear whether this is a single 
blob or it is composed of individual knots each of them associated 
to one of the linear filaments.

The curved jet-like system has a characteristic S-shape and marked 
point-symmetric brightness emission.  
The northern component (NCJ) seems to arise at a radial distance 
$\sim$17\arcsec\ and P.A.$\sim$35\degr.  
Then it steadily increases its P.A.\ as it proceeds farther, up to 
P.A.$\sim$55\degr\ at a distance $\sim$39\farcs3.  
As for the linear jet-like system, the emission in the [N~{\sc ii}] 
line brightens in the region farther than $\sim$30\arcsec\ from the 
nebular center.  
Indeed, only this bright [N~{\sc ii}] emission cap is detected 
for the southern component of the curved jet (SCJ).  
Both the southern and the northern curved jet-like features are knottier 
than their corresponding linear counterparts.

\subsubsection{Cometary knots}

A new morphological feature unveiled by the NOT narrow-band images 
is a group of cometary knots, most prominent in the [N~{\sc ii}] 
image (Figure~\ref{img2}).  
These low-excitation knots have bright, compact heads and tails 
up to 4\arcsec\ long that point outwards from the central region 
of the nebula.  
Some of them are projected onto the bipolar lobes, while some 
others are located outside the boundaries of the lobes.  
It is unclear whether all these knots have a connection with the 
jet-like systems, but the ``Northern Knot'' and ``Southern Knot'' 
in Figure~\ref{img2} seem to be the origin of the linear jets.

\subsubsection{Equatorial filamentary wisps}

Another interesting feature revealed by the NOT narrow-band images is a 
system of filamentary wisps protruding mostly from the equatorial regions 
of the nebula and along radial directions.  
These wisps are present in all the three narrow-band images, 
but especially in the [N~{\sc ii}] and H$\alpha$ images 
(Figures~\ref{img1} and \ref{img2}).  
As the angle between these wisps and the bipolar axis decreases, it becomes 
unclear whether the radial filamentary structures are wisps or cometary knot 
tails.  
Images at a higher angular resolution would be necessary to clearly 
distinguish wisps from cometary knot tails.

\begin{figure*}
\centering
\includegraphics[width=1.85\columnwidth]{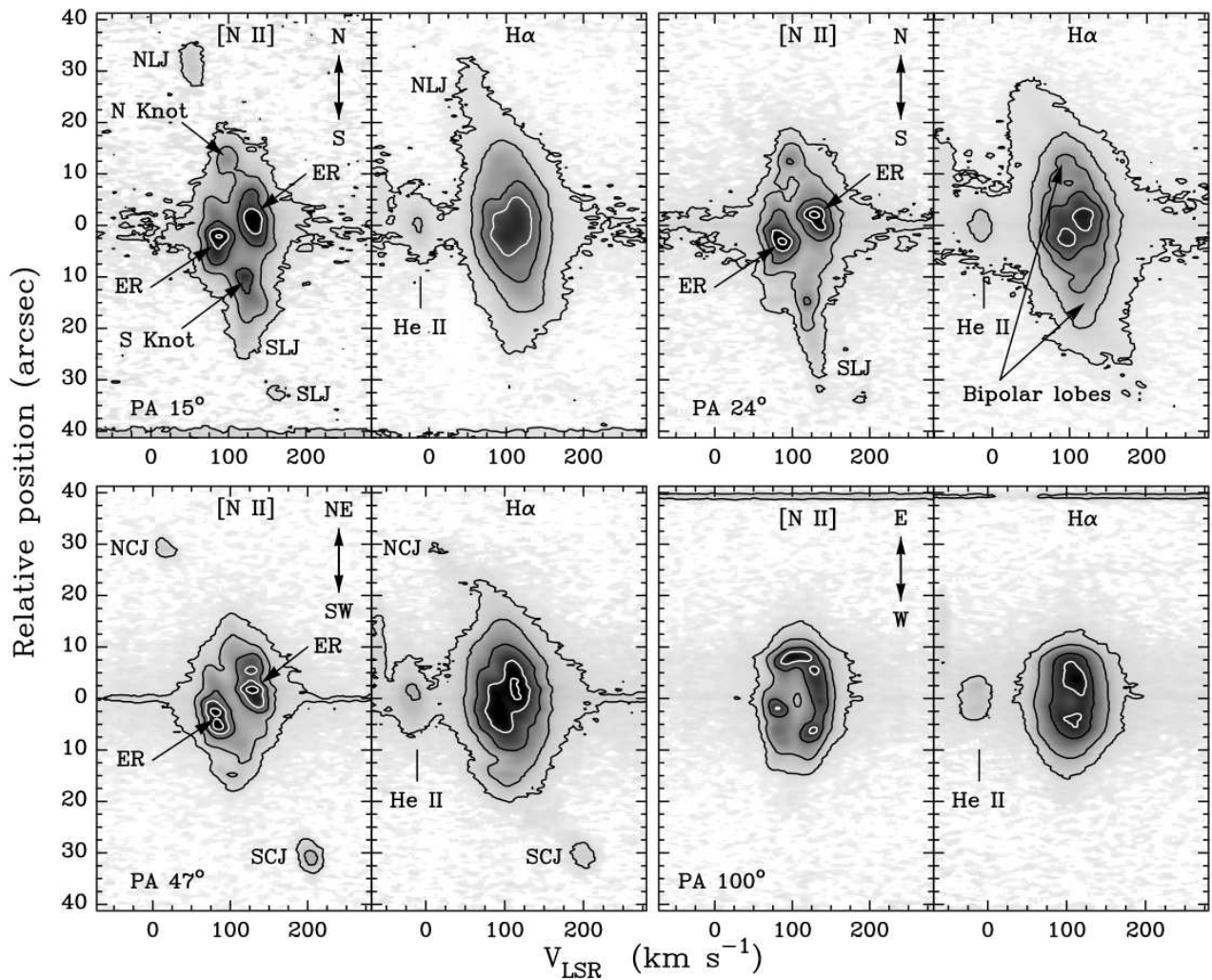}
\caption{
Gray-scale position-velocity (PV) echellograms of the H$\alpha$ and 
[N~{\sc ii}] emission lines of NGC\,6778 observed at four slits along 
the P.A.'s 15$^\circ$, 24$^\circ$, 47$^\circ$, and 100$^\circ$.  
Note the presence of emission from the He~{\sc ii} $\lambda$6560 line 
redwards of H$\alpha$.  
The morphological and kinematical features of interest are labeled 
on different echellograms: 
the Northern Linear Jet (NLJ), the Southern Linear Jet (SLJ), 
the Northern Curved Jet (NCJ), the Southern Curved Jet (SCJ), 
the Equatorial Ring (ER), 
the Bipolar Lobes, and 
the North Knot (N Knot) and South Knot (S Knot).  
The contour levels have been selected to emphasize the different 
features.  
}
\label{spec}
\end{figure*}

\subsection{Near-IR magnitudes of the central star}

The central star of NGC\,6778 has recently been recognized to be 
in a close binary system \citep{Miszalski_etal11}.  
This central star is detected in the broad-band near-IR images of 
NGC\,6778 as the relatively bright star marked by an arrow on the 
$J$ image in Figure~\ref{img_ir}.  
In order to derive the near-IR magnitudes of this star, we have compared 
its instrumental $JHK^\prime$ magnitudes with those of seven field stars 
in our images with $JHK_{\rm s}$ magnitudes in the 2MASS Point Source 
Catalogue \citep[PSC,][]{Sketal06}. 
After the appropriate conversion of $K^\prime$ and $K_{\rm s}$ magnitudes 
into $K$, this comparison yields $J = 16.87\pm0.12$, $H = 17.12\pm0.09$, 
and $K = 17.38\pm0.13$ for the central star, where the 1-$\sigma$ errors 
take into account the dispersion of the measurements of the comparison 
stars and the errors of the measured instrumental magnitudes. 
These magnitudes imply $J-H = -0.25\pm0.15$, $H-K = -0.26\pm0.16$, 
and $J-K = -0.51\pm0.18$ colors.  
Whereas the value of the $J-H$ color can be expected in hot WDs, we note 
that the colors derived using the $K$ band are anomalous.  
In particular, the value of the $J-K$ color is only marginally consistent 
with the values in the range from --0.20 to --0.30 mag expected for a WD 
\citep[][]{ZBM1991,GAN00}, and it seems to imply a lack of emission in the 
$K$ band (or an excess of emission in the $J$ band).  
We attribute this odd value to the difficulties in the measurement of the 
$K$ magnitude posed by the star faintness in this band and by the subtraction 
of the contribution of adjacent nebular emission.

The lack of near-IR excess and the observed $JHK$ magnitudes can be used to 
provide a crude constraint to the spectral type and luminosity class of the 
companion to the CSPN of NGC\,6778.  
A giant or sub-giant companion can be definitely discarded, but a 
dwarf companion is possible.  
The latest spectral type of this putative dwarf companion depends on the 
distance to the nebula: K0~V for a distance to the nebula $\leq$4 kpc, 
M0~V for a distance $\leq$2 kpc, and M5~V for a distance $\leq$1 kpc.  

% The derived magnitudes do not reveal any near-IR excess that could 
% be attributed to a late-type companion \citep[][]{ZBM1991,GAN00}.  
%
%  M5 V
%  K = 12.3 - 6.17 - 5 + 5*log(d)= 16.13 @ 1 kpc
%  K = ......................... = 17.64 @ 2 kpc
%  K = ......................... = 18.14 @ 4 kpc
%
%  M0 V
%  K = 8.8 - 3.65 - 5 + 5*log(d) = 15.15 @ 1 kpc
%  K = ......................... = 16.65 @ 2 kpc
%  K = ......................... = 18.15 @ 4 kpc
%
%  K0 V
%  K = 5.9 - 1.96 - 5 + 5*log(d) = 13.94 @ 1 kpc
%  K = ......................... = 15.44 @ 2 kpc
%  K = ......................... = 16.94 @ 4 kpc
%
%  G0 V
%  K = 4.4 - 1.41 - 5 + 5*log(d) = 13.00 @ 1 kpc
%  K = ......................... = 14.5  @ 2 kpc
%  K = ......................... = 16.00 @ 4 kpc

\subsection{Spatio-kinematical properties of NGC\,6778}

The preliminary description of the overall kinematics of NGC\,6778 
reported in Paper~I revealed a bipolar structure for the bright 
nebular shell, whereas the two pairs of jet-like features have large 
expansion velocities that increase outwards, i.e., they are real fast 
collimated outflows.  
Here we revisit these H$\alpha$ and [N~{\sc ii}] $\lambda$6583 
echellograms (Figure~\ref{spec}) in the light of the improved 
spatial resolution of the NOT narrow-band images, placing special 
emphasis on the kinematics of individual spatio-kinematical features.

\subsubsection{Equatorial ring}

\begin{figure*}
\centering
\includegraphics[width=1.80\columnwidth]{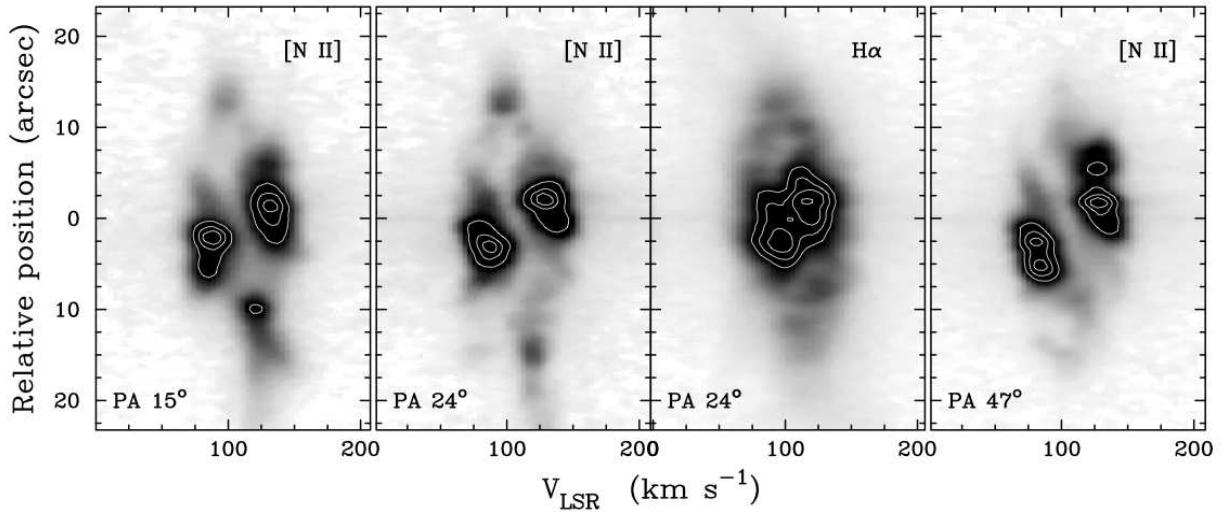}
\caption{
Grey-scale PV maps of H$\alpha$ at P.A.\ 24\degr, and [N~{\sc ii}] at 
P.A. \ 15\degr, \ 24\degr, and \ 47\degr, where the grey levels have 
been chosen to highlight the kinematics of the bipolar lobes.
}
\label{spec_bip}
\end{figure*}

First, we note that the echellograms of the [N~{\sc ii}] line from slits 
close to the symmetry axis (P.A.'s 15\degr, 24\degr, and 47\degr) show 
bright emission peaks detected at small radial distances.  
These emission peaks trace the emission from the distorted equatorial 
ring described in the images and have been consequently labeled as 
``ER'' in Figure~\ref{spec}.  
According to these echellograms, the northern regions of the equatorial 
ring recedes from us, while the southern regions moves towards us.  
The echellograms at P.A.'s 15\degr\ and 24\degr\ clearly 
reveal arc-like features departing from these peaks with 
increasing radial velocity.  
This can be interpreted as the opening of the ring or toroidal 
structure into the wider, higher velocity structure of the 
bipolar lobes, although it may alternatively be associated to 
the radial filaments of cometary knots and/or equatorial wisps.

The [N~{\sc ii}] echellogram of the slit at P.A.\ 100\degr\ reveals a 
donut-shaped line that is indicative of the expansion of the equatorial 
ring\footnote{
The higher surface brightness of the blue-shifted (northern) section 
of this line is most likely due to the slit position passing closer 
to the northern section of the ring.}, 
while the H$\alpha$ echellogram shows a rather filled line with two 
peaks closer than the edges of the [N~{\sc ii}] line.  
The velocity split derived from the [N~{\sc ii}] position-velocity map 
(52.2~km~s$^{-1}$) would imply an expansion velocity $\sim$26~km~s$^{-1}$, 
although it must be noted that the line-shape is not exactly an ellipse.  
These distortions of the velocity field are confirmed by the 
velocity split derived for the ring from the echellograms at 
the other P.A.'s: 43.5~km~s$^{-1}$ at P.A.\ 15\degr, 40.6~km~s$^{-1}$ 
at P.A.\ 24\degr, and 49.3~km~s$^{-1}$ at P.A.\ 47\degr.  
In view of the distorted morphology and kinematics of the ring, we must 
admit that the inclination angle of 15$^\circ$ derived in Paper~I under 
the assumption of a circular cross-section is questionable.

\subsubsection{Bipolar lobes}

Next we note the bipolar lobes, best seen in the H$\alpha$ echellograms 
along P.A.'s 15\degr\ and 24\degr\ since they are fainter in the 
low-ionization [N~{\sc ii}] line.  
The bipolar axis is tilted such that the north lobe is blue-shifted, 
conforming to the inclination of the equatorial ring.  
The difference in velocity between the tips of the 
north and south lobes is $\sim$20~km~s$^{-1}$.  
Since the velocity split of the equatorial ring is larger 
than the velocity difference between the bipolar tips, it 
indicates that the inclination angle of the bipolar axis 
with respect to the plane of the sky is small.  
If we assume a similar expansion, i.e., the expansion velocity increases with 
radial distance, then we can derive an inclination angle $\sim$12$^\circ$ and 
a deprojected expansion velocity at the tip of the bipolar lobes of 
50~km~s$^{-1}$.

A close inspection of the internal kinematics of the bipolar lobes 
(Figure~\ref{spec_bip}) reveals that it differs from the typical 
hour-glass-like expansion observed in other bipolar PNe 
\citep[e.g., K\,4-55;][]{GMS-R96}.  
Most notably, the [N~{\sc ii}] PV maps at P.A.\ 15\degr\ and 24\degr\ show 
a steady increase of the velocity split between the front and rear walls of 
the bipolar lobes up to a distance $\sim$7\arcsec, but, at larger angular 
distances, the velocity split abruptly decreases to rapidly converge 
into the N and S knots. 
Moreover, both the [N~{\sc ii}] and H$\alpha$ PV maps at these P.A.'s 
show the emission from the bipolar lobes delineating a twisted cavity 
(Figure~\ref{spec_bip}).

There is also evidence that implies that the bipolar lobes have been 
bored  along selected directions.  
At P.A.\ 24\degr, the front side of the southwestern lobe shows a 
breaking in the emission line at the position of a knot located at 
15\arcsec\ and $+$28~km~s$^{-1}$, i.e., the same projected distance as 
the S knot but at a different radial velocity.  
The PV map of the [N~{\sc ii}] line at P.A.\ 47\degr\ also shows 
openings at its tips, as well as a hook-shaped structure southwards 
between 9\arcsec\ and 18\arcsec\ and radial velocities between 
$+$17~km~s$^{-1}$ and $-$20~km~s$^{-1}$ that is hard to reconcile 
with an hour-glass-like shell.  
These results suggest that, similarly to the equatorial region, 
the bipolar lobes are disrupted and their kinematics distorted.  
The fact that the images show morphologically well defined lobes, while 
their kinematics is highly complex, may suggest that the disruption of 
the lobes is a relatively recent phenomenon that occurred on a time-scale 
shorter than the process that formed the bipolar shell.

\subsubsection{Linear and curved jets}

The echellograms displayed in Figure~\ref{spec} clearly confirm that 
the two jet-like systems described in \S3.1 are associated with high
velocity features.  
The radial velocity of the linear jets increases steadily with the 
distance to the central star of NGC\,6778 up to a maximum velocity 
split $\Delta v = 113\,\mathrm{km\,s}^{-1}$.  
Similarly, the curved jets show a velocity split 
$\Delta v = 192\,\mathrm{km\,s}^{-1}$.  
In both cases, the FWHM of the lines are very narrow, $\sim$23~km\,s$^{-1}$ 
at the tip of NLJ, and $\sim$16~km\,s$^{-1}$ for the tips of SCJ and NCJ.  
If we assume an inclination angle of 12$^\circ$ with respect to the plane 
of the sky, as for the bipolar lobes, then the deprojected velocities are 
270~km\,s$^{-1}$ for the linear jet, and 460~km\,s$^{-1}$ for the curved 
jet.  
We note that the assumption that these collimated outflows share 
the same inclination as the bipolar shell is questionable, very 
especially for the curved jet.  
At any rate, the jet-like morphology of these features, their large 
systemic velocities, and their small dispersion in velocity make us 
conclude that they are true fast collimated outflows that deserve 
to be named as jet.

\begin{figure}
\centering
\includegraphics[width=0.99\columnwidth]{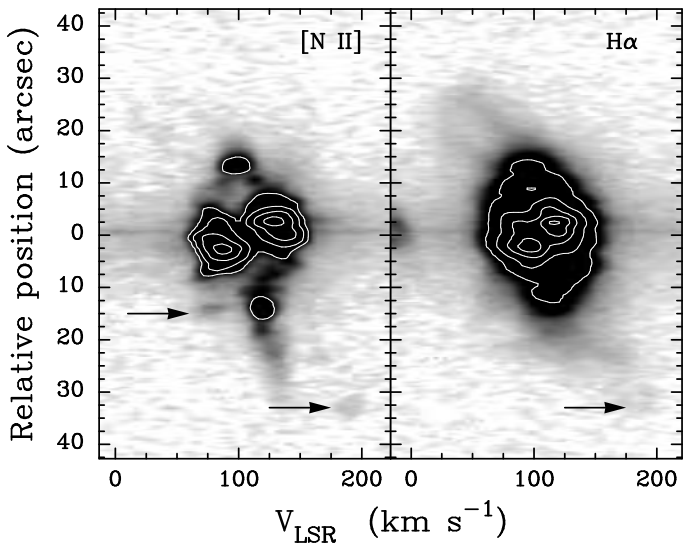}
\caption{
Grey-scale PV maps of the H$\alpha$ and [N~{\sc ii}] emissions at 
P.A.\ 24\degr. 
The arrows mark the location of knots described in Sect.\ 3.3.3.
}
\label{spec_pa24}
\end{figure}

We note that no expansion could be registered between the 2005 August 
and 2008 August images.  
Assuming conservatively that it would have been possible to measure a 
position shift $\sim$20\% the spatial resolution, i.e., $\sim$0\farcs16, 
the expansion velocity of the linear jet along the plane of the sky 
sets a lower limit for the distance to NGC\,6778 of $\ge$1 kpc.

The kinematics of the faintest features associated to these 
jets is best illustrated in Figure~\ref{spec_pa24}.  
The PV map of the [N~{\sc ii}] line at P.A.\ 24\degr\ shows a faint knot, 
coincident with the blob at the head of the SLJ, $\simeq$35$''$ apart 
from the central star, that exhibits a radial velocity with respect to 
the system of $+$83~km\,s$^{-1}$, much larger than the radial velocity at 
the tip of the jet, $\simeq$56~km\,s$^{-1}$.  
If this knot moved at an angle of 12\degr\ with respect to 
the plane of the sky, its velocity would be 400~km\,s$^{-1}$, 
i.e., larger than the velocity derived for LJ.  
This behavior may be indicative of an acceleration of SLJ at large 
distances, although the abrupt velocity increase results more likely 
from a change of the direction of the outflow that would be pointing 
more closely to the line of sight at this location.  
In the latter case, the abrupt change in direction could be tracing 
the interaction of SLJ with an ambient clump.  
We note that this phenomenon is not observed in the NLJ and CJ.

The complex kinematics in the region of LJ is best seen in the PV map of 
the H$\alpha$ line at P.A.\ 24\degr\ (Figure~\ref{spec_pa24}).  
The H$\alpha$ emission shows a bubble-like structure surrounding 
the emission from the NLJ.  
This bubble extends from $\simeq$12$''$ up to $\simeq$30$''$, is mainly 
shifted towards the blue (as it is NLJ), and presents a maximum velocity 
width $\simeq$100 km/s.  
A similar structure might be associated to SLJ, as suggested by the 
large velocity width ($\simeq$110~km\,s$^{-1}$) of the preferentially 
red-shifted, faint emission found at $\simeq$24$''$ from the center.  
Similar structures are not associated to the CJ. 
However, while the [N~{\sc ii}] emission is found mostly at the northern 
and southern heads of CJ, faint H$\alpha$ emission connects the heads of 
CJ with the outer regions of the bipolar lobes (see P.A.\ 47\degr\ in 
Figure~\ref{spec}). 
This faint H$\alpha$ emission reveals an outwards increase of 
the radial velocity along the CJ.

\section{Discussion}

The observational results presented in the previous section provide 
compelling evidence of strong dynamic interactions affecting the 
different nebular components of NGC\,6778.  
As for the case of NGC\,6326 described by \citet{Miszalski_etal11}, 
the equatorial ring of NGC\,6778 exhibits a highly distorted shape 
and morphological and kinematical signs of disruption. 
The equatorial wisps and cometary knots are also indicative of 
material being eroded from these knots and blown away.  
A fast stellar wind, a sudden mass ejection event, or the ablation 
by UV photons of dense knots may be responsible for the observed 
morphological features.

Similarly, the complex kinematics of the bipolar lobes suggests a 
dynamical interaction that has distorted its kinematics. 
There are interesting similarities with NGC\,6572 \citep{Miranda_etal99b} 
whose bipolar lobes also present a complex kinematics that cannot be 
interpreted with a simple (ellipsoidal o bipolar) geometry.  
It is also worth to remark the connection between the collimated outflows 
of NGC\,6778 and some bright knots at the tips of the bipolar shell, as 
seen in other PNe \citep[e.g., NGC\,6891;][]{Getal_00}.  
Very notably, the SLJ component seems to originate from a bright knot which is 
located in the PV plots at the point where the blue and red components of the 
bipolar lobes merge.  
The connection between the jets and knots at the tips of the bipolar 
lobes suggests either that the collimation of the outflows occurred 
at the tip of the bipolar lobes 
\citep[as in the converging flows scenario envisaged by][]{FBL96} 
or that the interaction of the fast collimated outflows have shaped 
the bipolar shell \citep[as suggested by][]{ST98}.  
The relatively undisturbed morphology of the bipolar lobes, in contrast with 
its kinematical properties, suggests that the bipolar lobes have only 
recently been shocked by the collimated outflows, i.e., that the bipolar 
lobes are older than the collimated outflows, formed in a late ejection 
event. 
The hole-like features of the bipolar lobes on the PV map at P.A.\ 
47$^{\circ}$ (Figure~\ref{spec_bip}), the direction along which the 
curved jets are detected, provides strong support for a lobe-jet 
interaction.

In this context, a comparison between the kinematical ages of shells and 
collimated outflows may be of interest to determine the time line of the 
formation of NGC\,6778, although the many dynamical effects involved 
and the uncertainty in their inclination can affect appreciable their 
absolute values \citep[][]{Corradi_etal99}.  
We derive a kinematical age 
$\sim$1700$\times{d}$ yr for the equatorial ring, 
$\sim$1600$\times{d}$ yr for the bipolar lobes, and 
 $\sim$650$\times{d}$ yr for the linear jet, 
where $d$ is the distance in kpc.  
% 
% ring:     4750 x  9.5 /  26 = 1735 yr 
% lobes:    4750 x 17   / 100 =  810 yr   /50 = 1610 yr
% lin. jet: 4750 x 36.5 / 265 =  650 yr  
%
The similar kinematical ages of the equatorial ring and bipolar lobes 
can be expected if they formed close in time.  
If we assume that the equatorial ring and nebular shell are indeed coeval and 
formed some 1700$\times{d}$ yr ago, then it can be envisaged that a more 
recent, $\sim$650$\times{d}$ yr old, series of collimated outflows interacted 
with the nebular shell.

The increasing expansion velocity of the collimated outflows with 
radius also suggests a free expansion that can be associated with 
a unique and sudden ejection event.  
There are several cases of this type of explosive-like, late events in 
PNe, especially associated to the late helium-shell flash associated to 
the born-again class of PNe \citep{GM96,Wesson_etal08} or to 
oxygen-neon-magnesium nova-like events suggested for the born-again 
PN A\,58 \citep{LdML11} or for BD+30\degr3639 \citep{Maness_etal03}.  
The chemical composition of NGC\,6778 is, however, typical of type~I bipolar 
PNe \citep{Milingo_etal10}, with no indications of neon overabundances or 
hydrogen-poor inclusions, although we note that these chemical abundances 
most likely correspond to the brightest regions of the main nebula.

Finally, we note that the inclination angle of the equatorial ring of 
NGC\,6778 inferred in the previous section, $\sim$12\degr, is larger 
than the value of 5\degr\ adopted by \citet{Miszalski_etal11} to 
account for the eclipses observed in its central star lightcurve\footnote{
We believe there is a typo in the statement made by 
\citet{Miszalski_etal11}.
They quote Paper~I for the inclination of 85\degr\ with respect to the line 
of sight, whereas the value of the inclination angle with respect to the 
plane of the sky provided in that reference is 15\degr\ rather than 5\degr.  
}.  
If we assume a 0.6 $M_\odot$ central star and a 0.3 $M_\odot$ dwarf companion 
of spectral type M4, which is not discarded by the near-IR data presented 
in section 3.2, the short orbital period of $\sim$0.15 days would imply a 
semimajor axis of $\sim$0.005 AU.  
For such a small orbital separation, the inclination angle of 12\degr\ 
would still produce eclipses of the primary star. 
%
% 0.005 AU x tan(12) = 1.6E10 cm = 0.23 R_Sun 
% Note, 0.005 AU equals 1.075 R_Sun
%

\section{Summary}

We have analyzed the detailed morphology and kinematics of NGC\,6778, 
a PN whose central star is in a binary system with a short orbital 
period, $\sim$0.15 days.  
The main nebular shell consists of a disrupted equatorial ring 
and kinematically disturbed bipolar lobes.  
The idea that the chaotic morphology and kinematics of NGC\,6778 stems from 
a violent event of sudden mass ejection is reinforced by the presence of 
equatorial wisps and cometary knots.  
There is an intimate connection between the fast collimated outflows and 
several kinematical and morphological features of the bipolar lobes.  
All these facts suggest that NGC\,6778 may have experienced a late  
explosive event that produced fast collimated outflows whose interaction 
with the nebular envelope has had dramatic effects in the nebular shaping.  
There is, however, no additional evidence for a late helium shell 
flash or a nova-like explosion.

\begin{acknowledgements}
Part of this work was supported by projects AYA2008-01934 and 
AYA2011-29754-C03-02 of the Spanish MICINN (co-funded by FEDER funds). 
LFM acknowledges partial support from INCITE09\,E1R312096, INCITE09\,312191PR, 
and IN8458-2010/061 of Xunta de Galicia (co-funded by FEDER funds). 
We gratefully thank the group of support astronomers of the Instituto de 
Astrof\'\i sica de Canarias (IAC) for making us available the narrow-band 
filters used at the NOT observations.   
We also thanks Dr.\ G.\ Ramos-Larios for his assistance during the 2008 
August NOT run and for the preparation of the color picture of NGC\,6778.  

\end{acknowledgements}


\begin{thebibliography}{}

\bibitem[Balick(1987)]{Balick87} 
Balick, B.\ 1987, 
\aj, 94, 671 

\bibitem[Balick \& Frank(2002)]{BF02} 
Balick, B., \& Frank, A.\ 2002, 
\araa, 40, 439 

\bibitem[Blackman(2009)]{Blackman09} 
Blackman, E.~G.\ 2009, 
in IAU Symp.\ 259, Cosmic Magnetic Fields: From Planets, to Stars and Galaxies, 
ed.\ K.G.\ Strassmeier, A.G.\ Kosovichev \& J.E.\ Beckman, 35

\bibitem[Bond \& Livio(1990)]{BL90}
Bond, H.~E., \& Livio, M.\ 1990, 
\apj, 355, 568 

\bibitem[Borkowski et al.(1997)]{Borkowski_etal97}
Borkowski, K.~J., Blondin, J.~M., \& Harrington, J.~P.\ 1997, 
\apjl, 482, L97 

\bibitem[Bujarrabal et al.(2002)]{Bujarrabal_etal02} 
Bujarrabal, V., Alcolea, J., S{\'a}nchez Contreras, C., \& Sahai, R.\ 2002, 
\aap, 389, 271 

\bibitem[Cant\'o et al.(1988)]{Canto_etal88} 
Cant\'o, J., Tenorio-Tagle, G., \& Roz\'yczka, M.\ 1988, 
\aap, 192, 287 

\bibitem[Corradi et al.(1999)]{Corradi_etal99} 
Corradi, R.~L.~M., Perinotto, M., Villaver, E., Mampaso, A., \& 
Gon{\c c}alves, D.~R.\ 1999, 
\apj, 523, 721 

\bibitem[Durand et al.(1998)]{DAZ98} 
Durand, S., Acker, A., \& Zijlstra, A.\ 1998, 
\aaps, 132, 13 

\bibitem[Frank et al.(1996)]{FBL96} 
Frank, A., Balick, B., \& Livio, M.\ 1996, 
\apjl, 471, L53 

\bibitem[Garc\'\i a-Segura et al.(2005)]{GarciaS_etal05} 
Garc\'\i a-Segura, G., L\'opez, J.~A., \& Franco, J.\ 2005, 
\apj, 618, 919 

\bibitem[Green et al.(2000)]{GAN00}  
Green, P.~J., Ali, B., \& Napiwotzki, R.\ 2000, 
\apj, 540, 992 

\bibitem[Guerrero \& Manchado(1996)]{GM96} 
Guerrero, M.~A., \& Manchado, A.\ 1996, 
\apj, 472, 711 

\bibitem[Guerrero \& Manchado(1998)]{GM98} 
Guerrero, M.~A., \& Manchado, A.\ 1998, 
\apj, 508, 262 

\bibitem[Guerrero et al.(1996)]{GMS-R96} 
Guerrero, M.~A., Manchado, A., \& Serra-Ricart, M.\ 1996, 
\apj, 456, 651 

\bibitem[Guerrero et al.(2001)]{Guerrero_etal01} 
Guerrero, M.~A., Miranda, L.~F., Chu, Y.-H., Rodr\'\i guez, M., \& 
Williams, R.~M.\ 2001, 
\apj, 563, 883 

\bibitem[Guerrero et al.(2000)]{Getal_00} 
Guerrero, M.~A., Miranda, L.~F., Manchado, A., \& V{\'a}zquez, R.\ 2000, 
\mnras, 313, 1 

\bibitem[Hora et al.(1999)]{HLD99} 
Hora, J.~L., Latter, W.~B., \& Deutsch, L.~K.\ 1999, 
\apjs, 124, 195 

\bibitem[Kwok(1983)]{Kwok83}
Kwok, S.\ 1983, 
in IAU Symp.\ 103, Planetary Nebulae, ed.\ D.R.\ Flower (Cambridge: 
Cambridge Univ.\ Press), 293

\bibitem[Kwok \& Su(2005)]{KS05} 
Kwok, S., \& Su, K.~Y.~L.\ 2005, 
\apjl, 635, L49 

\bibitem[Lau et al.(2011)]{LdML11} 
Lau, H.~H.~B., De Marco, O., \& Liu, X.-W.\ 2011, 
\mnras, 410, 1870 

\bibitem[Livio(2000)]{Livio00} 
Livio, M.\ 2000, 
in ASP Conf.\ Ser., Vol.\ 199, 
Asymmetrical Planetary Nebulae II: From Origins to Microstructures, 
ed.\ J.H.\ Kastner, N.\ Soker, \& S.\ Rappaport, 243 

\bibitem[Maestro et al.(2004)]{MGM04} 
Maestro, V., Guerrero, M.A., \& Miranda, L.F.\ 2004, 
in ASP Conf.\ Ser., Vol.\ 313, 
Asymmetrical Planetary Nebulae III: Winds, Structure and the Thunderbird, 
ed.\ M.\ Meixner, J.H.\ Kastner, B. Balick, \& N. Soker, 127 (Paper~I)

\bibitem[Maness et al.(2003)]{Maness_etal03} 
Maness, H.~L., Vrtilek, S.~D., Kastner, J.~H., \& Soker, N.\ 2003, 
\apj, 589, 439 

\bibitem[Milingo et al.(2010)]{Milingo_etal10} 
Milingo, J.~B., Kwitter, K.~B., Henry, R.~B.~C., \& Souza, S.~P.\ 2010, 
\apj, 711, 619 

\bibitem[Miranda et al.(1999a)]{Miranda_etal99a} 
Miranda, L.~F., Guerrero, M.~A., \& Torrelles, J.~M.\ 1999a, 
\aj, 117, 1421 

\bibitem[Miranda et al.(2010)]{MR-LG10}
Miranda, L.~F., Ramos-Larios, G., \& Guerrero, M.~A.\ 2010, 
\pasa, 27, 180 (Paper~II)

\bibitem[Miranda et al.(1999b)]{Miranda_etal99b}
Miranda, L.~F., V\'azquez, R., Corradi, R.~L.~M., Guerrero, M.~A., 
L\'opez, J.~A., \& Torrelles, J.~M. \ 1999b, \apj, 520, 714

\bibitem[Miszalski(2012)]{Miszalski12}
Miszalski, B.\ 2012, 
in IAU Symp.\ 283, Planetary Nebulae, an Eye to the Future, 
ed.\ A.\ Manchado, D.\ Sch\"onberner, \& L.\ Stanghellini 
(Cambridge: Cambridge Univ. Press), in press

\bibitem[Miszalski et al.(2011)]{Miszalski_etal11} 
Miszalski, B., Jones, D., Rodr{\'{\i}}guez-Gil, P., Boffin, H.~M.~J., 
Corradi, R.~L.~M., \& Santander-Garc{\'{\i}}a, M.\ 2011, 
\aap, 531, A158

\bibitem[Morris(1987)]{Morris87} 
Morris, M.\ 1987, 
\pasp, 99, 1115 

\bibitem[Patriarchi \& Perinotto(1991)]{PP91} 
Patriarchi, P., \& Perinotto, M.\ 1991, 
\aaps, 91, 325 

\bibitem[Perinotto et al.(2004)]{Perinotto_etal04} 
Perinotto, M., Patriarchi, P., Balick, B., \& Corradi, R.~L.~M.\ 2004, 
\aap, 422, 963 

\bibitem[Riera et al.(2003)]{Riera_etal03}
Riera, A., Garc{\'{\i}}a-Lario, P., Manchado, A., Bobrowsky, M., \& 
Estalella, R.\ 2003, 
\aap, 401, 1039 

\bibitem[Sahai(2000)]{Sahai00} 
Sahai, R.\ 2000, 
\apjl, 537, L43 

\bibitem[Sahai \& Nyman(2000)]{SN00} 
Sahai, R., \& Nyman, L.-{\AA}.\ 2000, 
\apjl, 538, L145 

\bibitem[Sahai \& Trauger(1998)]{ST98} 
Sahai, R., \& Trauger, J.~T.\ 1998, \aj, 116, 1357 

\bibitem[Schwarz et al.(1992)]{SCM92}
Schwarz, H.E., Corradi, R.L.M., \& Melnick, J.\ 1992, 
\aaps, 96, 23

\bibitem[Skrutskie et al.(2006)]{Sketal06} 
Skrutskie, M.~F., Cutri, R.~M., Stiening, R., et al.\ 2006, 
\aj, 131, 1163 

\bibitem[Soker(2004)]{Soker04} 
Soker, N.\ 2004, 
in ASP Conf.\ Ser., Vol.\ 313, 
Asymmetrical Planetary Nebulae III: Winds, Structure and the Thunderbird, 
ed.\ M.\ Meixner, J.H.\ Kastner, B. Balick, \& N. Soker, 562

\bibitem[Wesson et al.(2008)]{Wesson_etal08} 
Wesson, R., Barlow, M.~J., Liu, X.-W., Storey, P.~J., Ercolano, B., \& 
de Marco, O.\ 2008, 
\mnras, 383, 1639 

\bibitem[Zuckerman et al.(1991)]{ZBM1991} 
Zuckerman, B., Becklin, E.~E., \& McLean, I.~S.\ 1991, 
in ASP Conf.\ Ser., Vol.\ 14, 
Astrophysics with Infrared Arrays, 
ed.\ R.\ Elston, 161

\end{thebibliography}
\end{document}